\documentclass[prl,twocolumn,superscriptaddress]{revtex4}
\usepackage{empheq}
\usepackage{amsmath}
\usepackage{amssymb}
\usepackage{dsfont}
\usepackage{graphicx}
\usepackage{hyperref}
\usepackage{stackrel}
\begin{document}
\title{Phases of higher spin black holes: Hawking-Page, transitions between black holes and a critical point}
\author{M\'aximo Ba\~nados}
\email{maxbanados@fis.puc.cl}
\affiliation{Instituto de F\'isica, Pontificia Universidad Cat\'olica de Chile, \\
Casilla 306, Santiago, Chile}
\author{Gustavo D\"uring}
\email{gduring@fis.puc.cl}
\affiliation{Instituto de F\'isica, Pontificia Universidad Cat\'olica de Chile, \\
Casilla 306, Santiago, Chile}
\author{Alberto Faraggi}
\email{alberto.faraggi@unab.cl}
\affiliation{Instituto de F\'isica, Pontificia Universidad Cat\'olica de Chile, \\
Casilla 306, Santiago, Chile}
\affiliation{Departamento de Ciencias F\'isicas,Facultad de Ciencias Exactas, Universidad Andr\'es Bello, \\
Sazie 2212, Piso 7, Santiago, Chile}
\author{Ignacio A. Reyes}
\email{iareyes@uc.cl}
\affiliation{Instituto de F\'isica, Pontificia Universidad Cat\'olica de Chile, \\
Casilla 306, Santiago, Chile}
\affiliation{Insitut f{\"u}r Theoretische Physik und Astrophysik, Julius-Maximilians-Universit{\"a}t W{\"u}rzburg, \\ Am Hubland, 97074, Germany}

\begin{abstract}
We study the thermodynamic phase diagram of three-dimensional $sl(N;\mathds{R})$ higher spin black holes. By analyzing the semi-classical partition function we uncover a rich structure that includes Hawking-Page transitions to the AdS$_3$ vacuum, first order phase transitions among black hole states, and a second order critical point. Our analysis is explicit for $N=4$ but we extrapolate some of our conclusions to arbitrary $N$. In particular, we argue that even $N$ is stable in the ensemble under consideration but odd $N$ is not.
\end{abstract}

\maketitle

Soon after black hole entropy \cite{bekenstein1973black,bardeen1973four} and radiation \cite{hawking1975particle} were discovered, Gibbons and Hawking \cite{gibbons1977action} showed that these properties can be derived directly from the Euclidean gravitational action. Black holes are now understood as part of a thermodynamical system with an associated semi-classical partition function
\begin{empheq}{alignat=7}\label{Z0}
	Z(\beta)&=\sum_Me^{-\beta M+S(M)}\,.
\end{empheq}
For Schwarzschild black holes the entropy takes the famous value $S=\textrm{Area}/(4G)=4\pi GM^2$ \footnote{We work in natural units $\hbar=c=k_B=1$.}.

This partition function can be extended to more general black holes in various dimensions. Of particular interest in recent years has been a relatively new family of configurations, namely, three-dimensional black holes carrying higher spin charges \cite{gutperle2011higher}.

In this paper we study the thermodynamics of higher spin theories by emphasizing the role of the partition function \eqref{Z0}. We uncover a rich structure with several interesting features: i) the existence of Hawking-Page transitions from black holes to the $AdS_3$ background; ii) discontinuous phase transitions among black hole states with different macroscopic properties (van der Waals-like); and iii) a second order transition and a critical point. For related work see \cite{David:2012iu,Chen:2012ba,Chowdhury:2013roa,Ferlaino:2013vga}.

We start by reviewing some applications of (\ref{Z0}). The partition function is dominated by the configuration that minimizes the action
\begin{empheq}{alignat=7}\label{F0}
	\Gamma_\beta(M)&=\beta M-S(M)\,.
\end{empheq}
In a more general setup, $\Gamma$ will depend on additional charges and chemical potentials. This function, related to the mean field free energy, encodes the thermodynamic structure of the gravitational system and will be our main tool to analyze its phases. Consider Schwarzschild black holes, for example. A quick look shows that $\Gamma_{\beta}(M)$ does not have a minimum at all; the value $M=\beta/(8\pi G)$ is a maximum, revealing the well-known instability of this system in the canonical ensemble.

The instability of Schwarzschild black holes can be cured either by putting the system in a box \cite{Whiting:1988qr} or by adding a negative cosmological constant \cite{Hawking:1982dh}, the latter case leading to the celebrated Hawking-Page transition. It is instructive to understand this phenomenon from the point of view of the action \eqref{F0}.
 
\begin{figure}[h]
\fbox{\includegraphics[width=0.2251\textwidth]{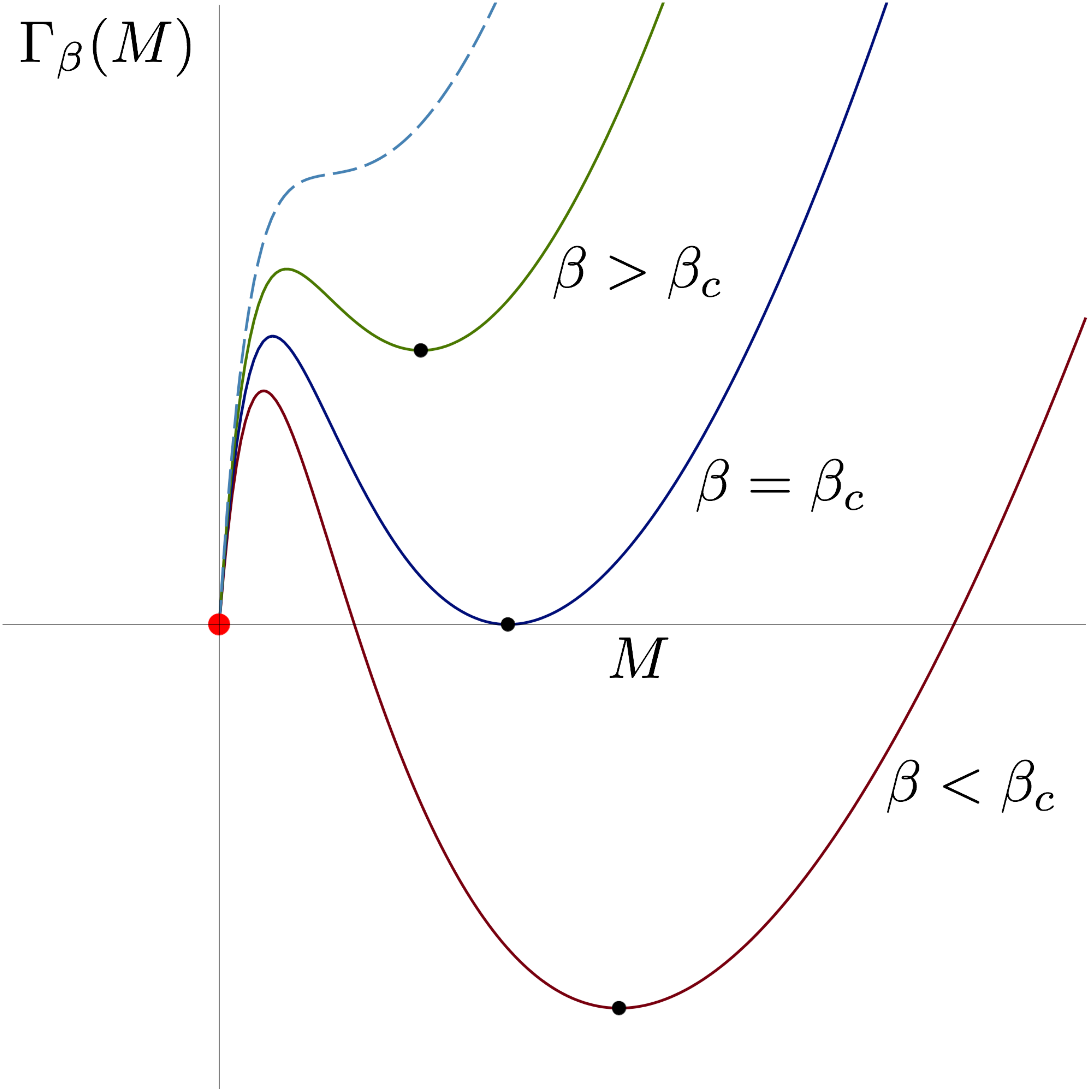}}
\fbox{\includegraphics[width=0.2251\textwidth]{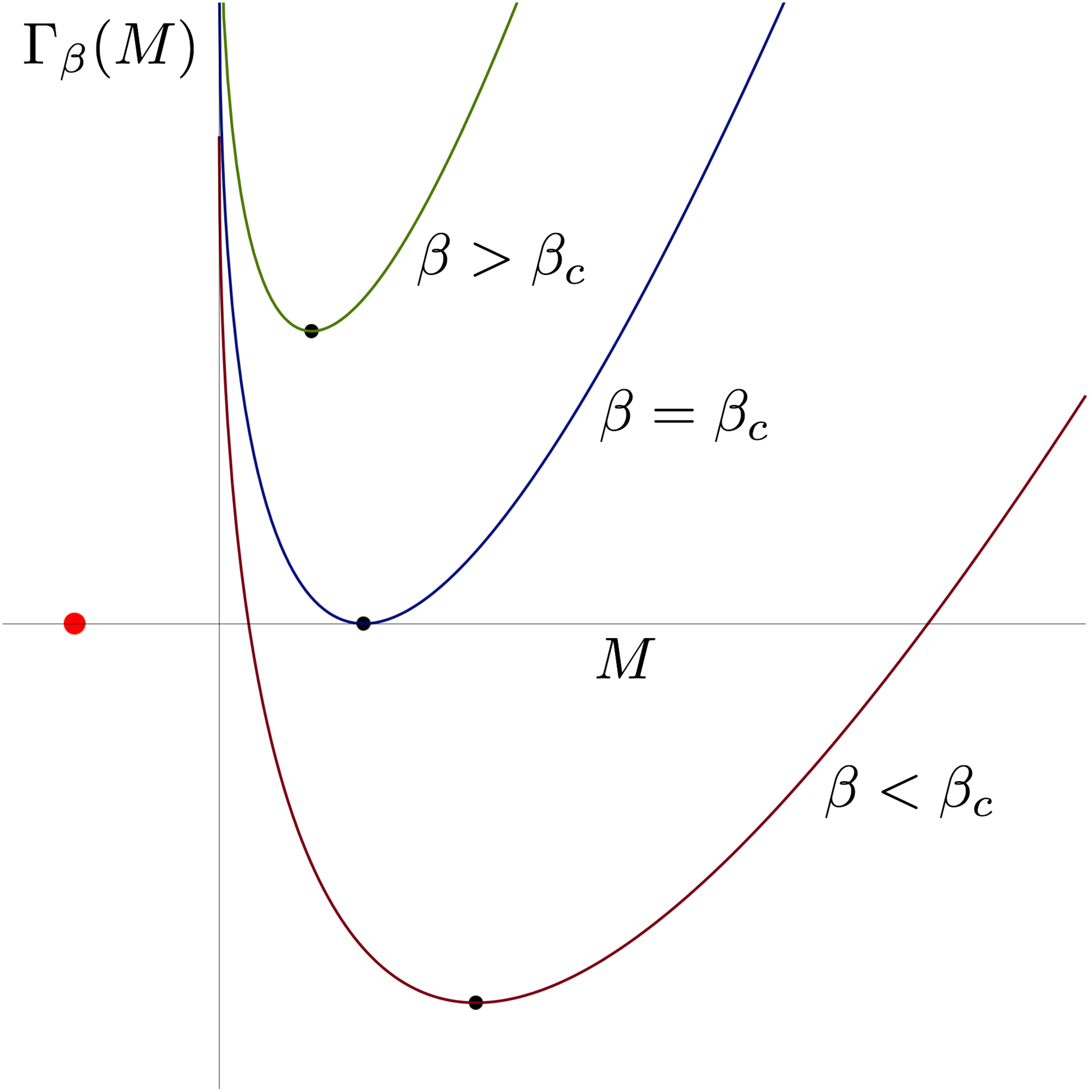}}
\caption{Hawking-Page transition in $3+1$ Schwarzschild-AdS black holes (left) and $2+1$ BTZ black holes (right). The red dots represent the $AdS_4$ ($M=0$) and $AdS_3$ ($M=-1/(8G)$) ground states, respectively. The dashed curve has no local minima but plays no role in the analysis.}
\label{HP}
\end{figure}

In Figure \ref{HP} (left) we plot $\Gamma_\beta(M)$ for $3+1$ Schwarzschild-AdS black holes for different values of $\beta$ \footnote{The entropy of $3+1$ Schwarzschild-AdS black holes is $S(M)=\pi r_+^2(M)/G$, where $r_+(M)$ is the unique real solution to the equation $1-2GM/r_++r_+^2/\ell^2=0$. In particular, $S(0)=0$; the AdS$_4$ vacuum has no entropy. The critical temperature is $\beta_c=\pi\ell$.}. At high temperatures (burgundy curve, small $\beta$), there is a clear minimum satisfying
\begin{equation}\label{m}
\beta = {\partial S(M) \over  \partial M}. 
\end{equation}
As the temperature drops, however, the solution to this equation ceases to be the global minimum of $\Gamma_{\beta}(M)$. For $\beta > \beta_c$, the AdS$_4$ background with $M=0$ is the preferred state. This transition from the black hole dominated phase to the vacuum is called Hawking-Page transition. Notice that the minimum of $\Gamma_{\beta}(M)$ is continuous at $\beta=\beta_c$ but its derivative is not.

The analysis for three-dimensional BTZ black holes is similar, except that the AdS vacuum is now a bound state separated from the black hole continuum by a mass gap. In Figure \ref{HP} (right) we plot $\Gamma_{\beta}(M)=\beta(M-M_{AdS})-S(M)$ for three values of $\beta$ \footnote{The entropy of $2+1$ BTZ black holes is $S(M)=\ell\pi\sqrt{2M/G}$. In this normalization the AdS$_3$ vacuum corresponds to $M=-1/(8G)$. Of course, it carries no entropy. The critical temperature is $\beta_c=2\pi\ell$.}. For convenience we have shifted the action such that $\Gamma_{\beta}(M_{AdS})=0$. We see that a local minimum satisfying \eqref{m} exists for all temperatures but only for $\beta<\beta_c$ does this state have less action than AdS$_3$ space.

We now move on to study the phase structure of three-dimensional higher spin gravity. This theory is topological in nature and does not have a known description in terms of metric fields ($g_{\mu\nu}$, $g_{\mu\nu\rho}$, etc.). Instead, it must be formulated as a Chern-Simons theory where the basic variables are $sl(N;\mathds{R})$ ($N\times N$, real, traceless) matrices $A_{\mu}$. In a radial gauge, and restricting to static and circularly symmetric configurations, one is left with two such matrices, $A_t$ and $A_\varphi$, satisfying
\begin{empheq}{alignat=7}\label{eqm}
	[A_t,A_\varphi]&=0\,.
\end{empheq}
This is the remnant of the Chern-Simons equations of motion. We refer the reader to the extensive literature on this subject for more details.

The gauge invariant information carried by the fields is characterized by the $N-1$ charges
\begin{empheq}{alignat=7}\label{casimir}
	Q_n&=\frac{1}{n}\textrm{Tr}\left(A_{\varphi}^n\right)\,,
	&\qquad
	n&=2\,,\ldots\,,N\,.
\end{empheq}
The possible values these charges can take depend on the spatial topology of spacetime. We shall consider two classes of solutions. First, there is the AdS$_3$ vacuum, for which the spatial topology is a plane and the cycle $\varphi\sim\varphi+2\pi$ is contractible. As a consequence, the eigenvalues of $A_\varphi$ are imaginary and quantized, so as to render a smooth field \footnote{The holonomy is $Pe^{\oint A^{AdS}_{\varphi}}=(-1)^{N-1}\mathds{1}_{N\times N}$.}:
\begin{empheq}{alignat=7}\label{adsphi}
	\textrm{Eigen}\left(A^{AdS}_{\varphi}\right)&=\frac{i}{2}\left(N-1,N-3,\ldots,1-N\right)\,.
\end{empheq}
The corresponding charges can be computed from \eqref{casimir}. Notice that $Q_n^{AdS}=0$ for odd $n$. For $N=4$, which will be our main example, we find 
\begin{equation}\label{adsq}
Q^{AdS}_2=-{5 \over 2}\,, \ \ \ \  Q^{AdS}_3=0\,, \ \ \ \  Q^{AdS}_4= {41 \over 16}\,.
\end{equation} 
The time direction is non-compact so the matrix $A^{AdS}_t$ is only restricted by \eqref{eqm} and not by regularity. This is important. The residual freedom in $A^{AdS}_t$ is just what is needed in order to match the chemical potentials of AdS$_3$ to those of black holes (just like thermal AdS and Schwarzchild-AdS can both be put at the same temperature). 

The second class of solutions we are interested in are black holes. The spatial topology in this sector is a semi-infinite cylinder, with the boundary at $r=0$ corresponding to the horizon. This space is homeomorphic to the punctured disk. Since the cycle $\varphi\sim\varphi+2\pi$ is not contractible, $A_{\varphi}$ (and therefore $Q_n$) is left unrestricted. We take the eigenvalues of $A_{\varphi}$ to be real for black hole configurations, in consistency with the definition given in \cite{BCFJ} in a supersymmetric context. For $N=4$ we parametrize the eigenvalues as
\begin{empheq}{alignat=7}\nonumber
	\textrm{Eigen}\left(A_{\varphi}\right)&=\frac{1}{2}\left(2\lambda_1+\lambda_2,2\lambda_1-\lambda_2,\right.
	\\
	&\left.-2\lambda_1+\lambda_3,-2\lambda_1-\lambda_3\right)\,.\label{aphi}
\end{empheq}
It follows that the black hole charges read
\begin{empheq}{align}
	Q_2&=2\lambda_1^2+\frac{1}{4}\lambda_2^2+\frac{1}{4}\lambda_3^2\,, \nonumber
	\\
	Q_3&=\frac{1}{2}\lambda_1\left(\lambda_2-\lambda_3\right)\left(\lambda_2+\lambda_3\right)\,, \nonumber
	\\
	Q_4&=\lambda_1^4+\frac{1}{32}\left(\lambda_2^4+\lambda_3^4\right)+\frac{3}{4}\lambda_1^2\left(\lambda_2^2+\lambda_3^2\right)\,. \label{Qn}
\end{empheq}
In this sector, the matrix $A_t$ is constrained by regularity to be \footnote{The absence of the factor of $i$ in \eqref{At} is due to the Euclidean continuation. It reappears in the holonomy $Pe^{i\oint A_t}=(-1)^{N-1}\mathds{1}_{N\times N}$.}
\begin{empheq}{alignat=7}\label{At}
	\textrm{Eigen}\left(A_t\right)&=\frac{1}{2}\left(N-1,N-3,\ldots,1-N\right)\,.
\end{empheq}
This is because the time cycle $t\sim t+2\pi$ is contractible in the black hole topology. Finally, black holes have an entropy given by \cite{banados2012action,Bunster:2014mua,Perez:2013xi,deBoer:2013gz}
\begin{empheq}{alignat=7}
	S(Q)&=\textrm{Tr}\left(A_tA_{\varphi}\right)\,.
\end{empheq}
For $N=4$ this yields
\begin{empheq}{alignat=7}
	S(\lambda)&=\frac{1}{2}\left(8\lambda_1+\lambda_2+\lambda_3\right)\,. \label{S}
\end{empheq}

We are now ready to display the grand canonical partition function \footnote{Higher spin gravity is actually described by two independent copies of the $sl(N)$ Chern-Simons action. Since the total partition function factorizes, we focus on one sector only.} we aim to calculate:
\begin{empheq}{alignat=7}
	Z(\mu_2,\mu_3,...\mu_n)&=1+\sum_{\left\{\substack{\textrm{\tiny{black}}\\\textrm{\tiny{holes}}}\right\}}e^{-k\Gamma_{\mu}(Q)}\,,\label{ZN}
\end{empheq}
where the action is
\begin{empheq}{alignat=7}\label{IN}
	\Gamma_{\mu}(Q)&=\sum_{n=2}^{N}\mu_n\left(Q_n-Q_n^{AdS}\right)-S(Q)\,.
\end{empheq}
The first term `1' in \eqref{ZN} corresponds to the AdS$_3$ bound state, whose contribution has been subtracted in \eqref{IN} so that $\Gamma_{\mu}(Q^{AdS})=0$. The AdS charges have fixed values (for $N=4$, see \eqref{adsq}). The sum is then taken over the spectrum of black holes, that is, over all charges consistent with real eigenvalues of $A_\varphi$. Also, for convenience we have factored out the coupling constant ``$k$" (Chern-Simons level). This parameter is related to the central charge by
\begin{empheq}{alignat*=7}
	k&=\frac{2\pi c}{N(N^2-1)}\,,
	&\qquad
	c&=\frac{3\ell}{2G}\,,
\end{empheq}
and is a measure of the number of degrees of freedom in the system. The steepest descent approximation is justified in the limit $k\sim c\rightarrow\infty$. The real parameters $\mu_{n}$ are chemical potentials conjugate to $Q_n$. See \cite{banados2012action} for details on this. The case of pure gravity is recovered for $N=2$ after identifying $\mu_2=\beta/(2\pi l)$ and $Q_2=2GM$.

We emphasize that the study of the partition function (\ref{ZN}) represents a well-posed problem on its own right, independent from its relation to Chern-Simons theories and higher spin black holes. In fact, this problem has striking similarities with the mean field description of some condensed matter systems such as liquid crystals \cite{prost1995physics,andrienko2006introduction,zheng2011maier}.

Notice that, for any $N$, the charges $Q_{2n}$ are always semi-positive while $Q_{2n+1}$ have no definite sign. This explains why even $N$ can yield a stable partition function but odd $N$ cannot. Indeed, the action \eqref{IN} is a polynomial of degree $N$ in the eigenvalues of $A_{\varphi}$. Since the leading terms enter as $\Gamma_{\mu}(\lambda)=\mu_NQ_N(\lambda)+\cdots$, the conditions $N\in2\mathds{N}$ and $\mu_N>0$ guarantee that $\Gamma_{\mu}(\lambda)$ is bounded from below. The sum \eqref{ZN} then converges. It turns out that $N=4$ is the simplest, non-trivial, stable theory; $N=2$ exhibits a Hawking-Page transition but no transitions between black hole states, and $N=3$ is unstable. From now on we concentrate on $N=4$.

In principle, the computation of (\ref{ZN}) involves a sum over the charges $Q_n$. This is inconvenient because the entropy \eqref{S} has a very simple form in terms of the eigenvalues but not as a function of the charges themselves. Expressing $S$ as a function of $Q_n$ would involve inverting \eqref{Qn}. Happily, this is not necessary. We shall now argue that the sum over charges can be traded for a sum over eigenvalues, up to logarithmic corrections that we discard in large $k$ limit. 

From (\ref{Qn}) we find that the Jacobian for the change of variables $Q_n  \rightarrow \lambda_i$ is
\begin{empheq}{alignat=7}\label{J}
	\nonumber
	\left|\frac{\partial Q}{\partial\lambda}\right|&=\frac{1}{32}\lambda_2\lambda_3\left(4\lambda_1+\lambda_2+\lambda_3\right)\left(4\lambda_1-\lambda_2+\lambda_3\right)
	\\
	&\times\left(4\lambda_1+\lambda_2-\lambda_3\right)\left(4\lambda_1-\lambda_2-\lambda_3\right)\,.
\end{empheq}
We see that $\left|{\partial Q \over \partial \lambda}\right| = 0$ happens precisely when two or more eigenvalues of $A_\varphi$ coincide. In that case, $A_\varphi$ is non-diagonalizable and the solution becomes extremal \cite{BCFJ}. Of course, we restrict our attention to solutions bounded by extremal black holes; the points where the map $Q_n \rightarrow \lambda_i$ is not invertible are never touched. For orientation, recall that in the spectrum of 2+1 black holes, the angular momentum $J$ is bounded by $-M < J < M$, with $J = \pm M$ corresponding to extremal solutions. The map $Q_n \rightarrow \lambda_i$ fails to be invertible at the extreme points $J=\pm M$. Going back to the general case, the domain of $\lambda_i$ should lie within any of the ``wedges'' defined by the planes $\left|{\partial Q \over \partial \lambda}\right|=0$. Different wedges correspond to different branches of the inverse map $Q\rightarrow\lambda(Q)$ and have different entropy functions $S(Q)$. We will work in the wedge that includes the BTZ black hole ($\lambda_1=\lambda_2=\lambda_3>0$), for which all the factors in \eqref{J} are positive.

So, given a set of values for the chemical potentials $\mu_2$, $\mu_3$ and $\mu_4$, we want to compute the values of $\lambda_1$, $\lambda_2$ and $\lambda_3$ that minimize the action $\Gamma_{\mu}(\lambda)$. This configuration will dominate the partition function \eqref{ZN}. In particular, we would like to study the continuity of the $\lambda_i$ obtained in this way as one varies $\mu_2$, $\mu_3$ and $\mu_4$. 

First, we have checked explicitly, within a wide range of chemical potentials, that the minimum of $\Gamma_{\mu}(\lambda)$ is never achieved by extremal black holes. The minimum either occurs for the $AdS_3$ groundstate, with ${\Gamma}_\mu(Q^{AdS})=0$, or it lies in the interior of the BTZ wedge \footnote{We have also checked that if one does not restrict the eigenvalues to any particular wedge, the global minimum of $\Gamma_{\mu}(\lambda)$ in the black hole sector is always found in the BTZ wedge and not in any other.}. We do not need to worry about extremal solutions. 

\begin{figure}[h]
	\includegraphics[width=0.4\textwidth]{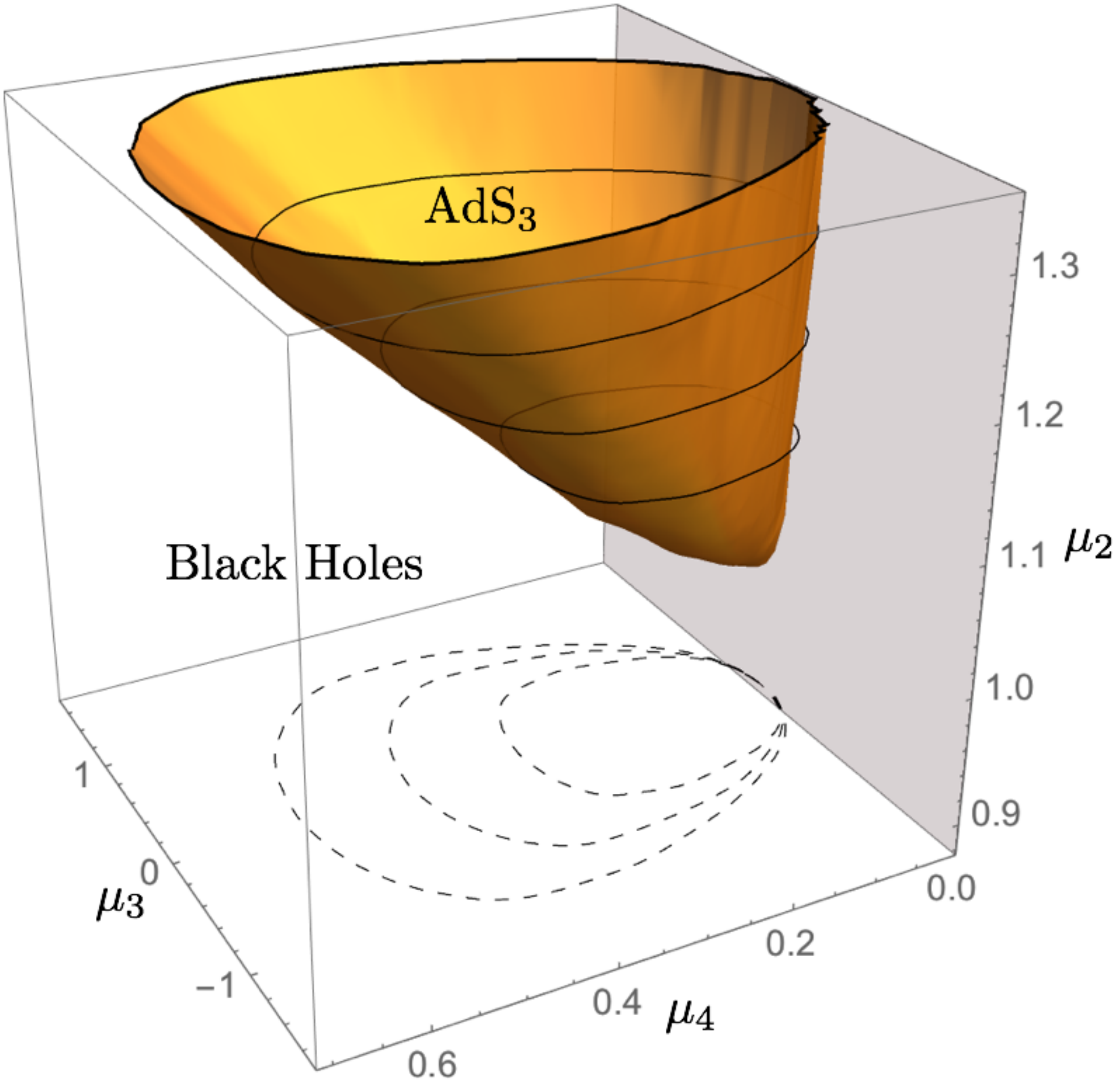}
	\caption{Hawking-Page surface for the $sl(4;\mathds{R})$ theory. The minimum is located at $\mu_2=1$, $\mu_3=0$, $\mu_4=0$. This coincides with the critical temperature of $2+1$ BTZ black holes after identifying $\mu_2=\beta/(2\pi\ell)$.}\label{HPHS}
\end{figure}

Next, we separate the region in the space of chemical potentials where black hole states dominate from the region where the AdS ground state is preferred. The interface between these two regions is defined by the equation
\begin{empheq}{alignat=7}\label{Def HPHS}
	\underset{\left\{\substack{\textrm{{\tiny black}}\\\textrm{{\tiny holes}}}\right\}}{\textrm{min}}\,\Gamma_{\mu}(Q)&=\Gamma_{\mu}\left(Q^{AdS}\right),
\end{empheq}
and its graphical representation is shown in Figure \ref{HPHS}.
The interior of the surface corresponds to the AdS dominated phase. Outside black holes dominate. Crossing this surface in any direction gives a first order Hawking-Page transition.

Let us now concentrate on the region of black hole dominance and look for the global minimum satisfying
\begin{empheq}{alignat=7}\label{min}
		\frac{\partial \Gamma_{\mu}(\lambda)}{\partial\lambda_i}&=0\,.
\end{empheq}
(Recall that $\Gamma_\mu(\lambda)$ is built from \eqref{IN} after using \eqref{Qn} to write the charges in terms of the eigenvalues, together with the expressions (\ref{S}) for the entropy and \eqref{adsq} for the background charges). These equations can be simplified by setting $\mu_2=5\mu$ and rescaling $\mu_3\rightarrow \mu_3 \mu^2$,  $\mu_4\rightarrow \mu_4 \mu^3$ and $\lambda_i\rightarrow \lambda_i/\mu$. The $\mu$ dependence then drops out from \eqref{min}, reducing the black hole thermodynamics to a two dimensional phase diagram. 

\begin{figure}[h]
    \centering
    \includegraphics[width=0.475\textwidth]{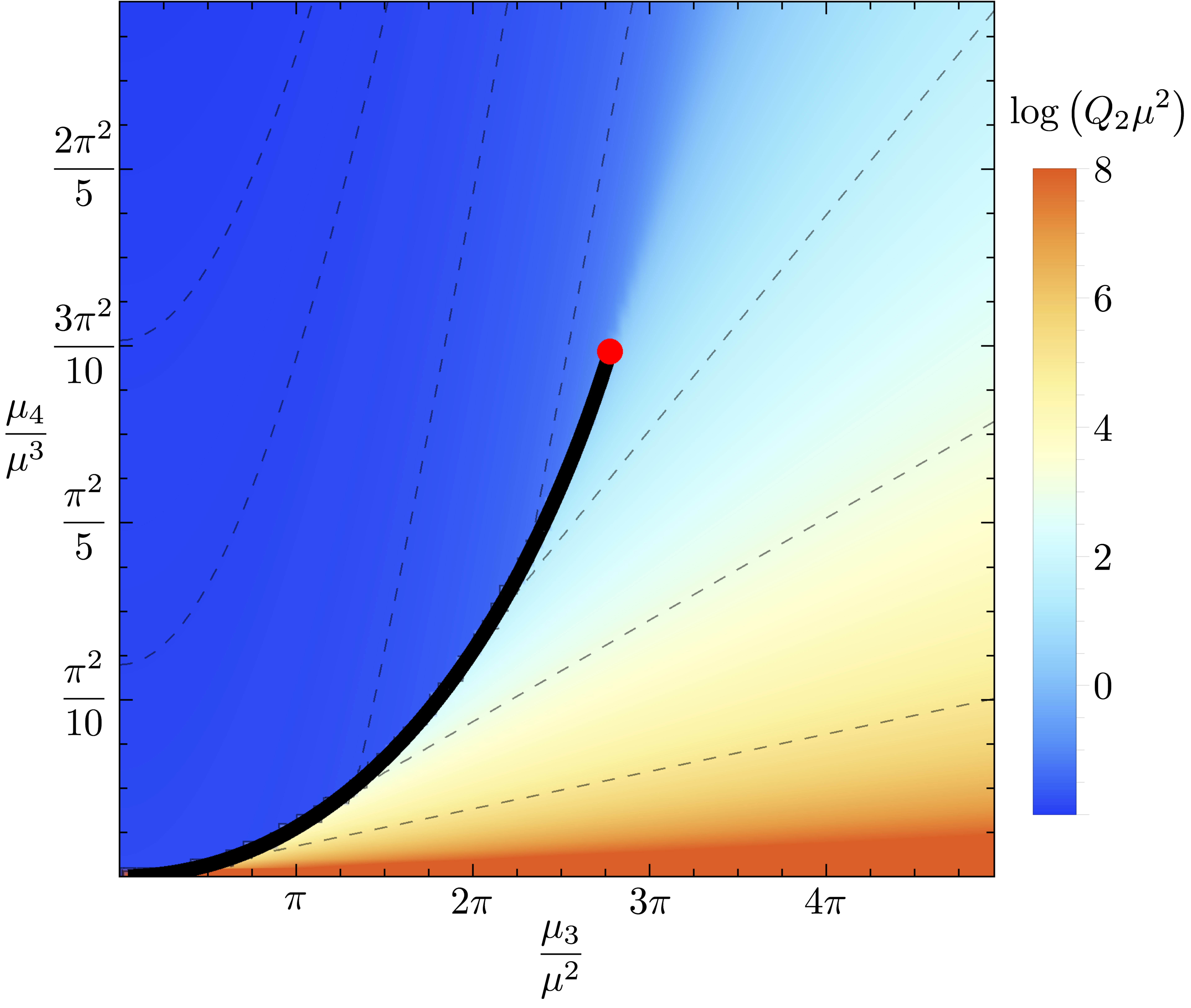}
    \caption{Phase diagram of the $N=4$ system, with the second order critical point in red and the first order critical line in black. The complete diagram is symmetric with respect to the vertical axis. $\mu=\mu_2/5$.}
    \label{fig:phase}
\end{figure}

Figure \ref{fig:phase} shows the phase diagram for the charge $Q_2$. 
Similar results are obtained for the other two charges. One observes that from the origin stems a critical line across which the system exhibits a first order phase transition between two macroscopically different black hole states. The line ends at a critical point (in red). We now show that at the critical point a second order phase transition takes place in which the minima of $\Gamma_{\mu}(\lambda)$ become degenerate.  

For systems with a single order parameter a critical point occurs when the first, second and third derivatives of the free energy vanish. The simplest generalization to the case with multiple order parameters is to demand that the Hessian matrix has one null eigenvalue, with the rest being strictly positive. Calling $v^i$ the corresponding normalized eigenvector, we further require that the third derivative of $\Gamma_{\mu}(\lambda)$ along $v^i$ vanishes. Thus, in addition to \eqref{min}, the  critical point must satisfy
\begin{empheq}{alignat=7}\label{crit}
	v^j{\partial^2\Gamma_\mu(\lambda) \over\partial \lambda_i \partial\lambda_j} &=0\,,
	&\quad
	v^iv^jv^k{\partial^3\Gamma_\mu(\lambda) \over \partial \lambda_i\partial\lambda_j \partial\lambda_k}&=0\,.
\end{empheq}
To ensure that we still have a minimum, the fourth derivative along $v^i$ should be positive. Following this approach we find that the $N=4$ higher spin theory exhibits two mirror critical points, the first of which is located at
\begin{empheq}{align}
	\nonumber
	\mu_2&=5\,\mu \,,
	&
	\lambda_1&=\frac{0.3299}{\mu}\,,
	&
	v^1&=-0.2061\,,
	\\\nonumber
	\mu_3&=8.7184\,\mu^2\,,
	&
	\lambda_2&=\frac{0.0854}{\mu}\,,
	&
	v^2&=0.0349\,,
	\\
	\mu_4&=2.9299\,\mu^3\,,
	&
	\lambda_3&=\frac{1.0259}{\mu}\,,
	&
	v^3&=-0.9779\,,
\end{empheq}
and corresponds to the one displayed in Figure \ref{fig:phase}. The second point (not shown in Figure \ref{fig:phase}) is obtained by exchanging $\lambda_2\leftrightarrow\lambda_3$ ($Q_3\rightarrow-Q_3$) and $\mu_3\rightarrow-\mu_3$, which is a symmetry of the action \eqref{IN}. The values of the charges and the entropy at the critical points can be computed directly from (\ref{Qn}) and (\ref{S}). One can check that in the range $1.0886<\mu<6.3591$ the second order phase transition takes place inside the Hawking-Page surface and is therefore not relevant. The endpoints of this interval correspond to the intersection of the critical point with the Hawking-Page surface.

\begin{figure}[h]
    \centering
    \includegraphics[width=0.475\textwidth]{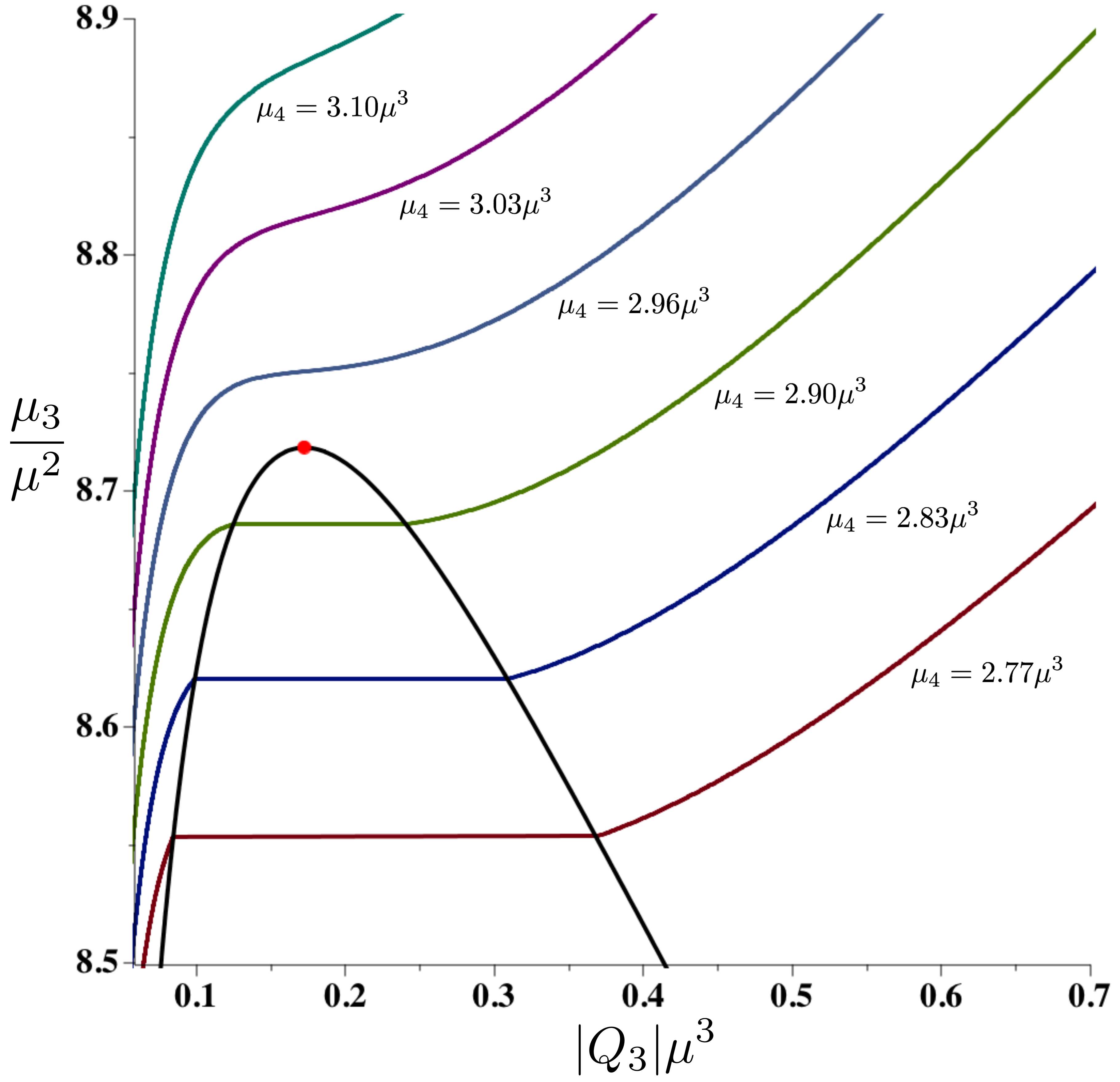}
    \caption{$\mu_3$ - $Q_3$ diagram with iso - $\mu_4$ curves.}
    \label{fig:VdW3}
\end{figure}

\begin{figure}[h]
    \centering
    \includegraphics[width=0.475\textwidth]{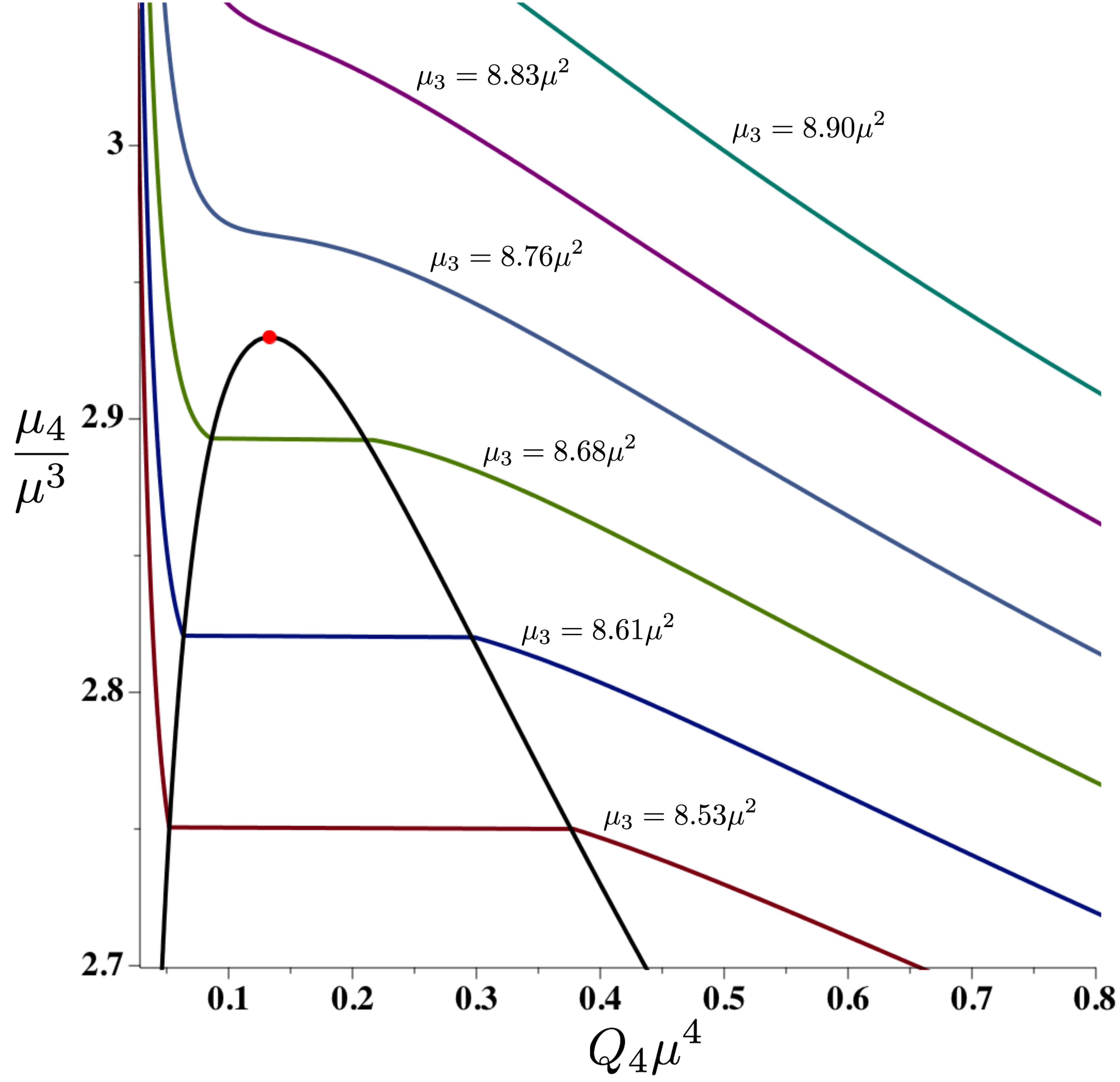}
    \caption{$\mu_4$ - $Q_4$ diagram with iso - $\mu_3$ curves.}
    \label{fig:VdW4}
\end{figure}

The phase diagram ``$\mu_3$-$\mu_4$" is qualitatively similar to a ``P-T" diagram for a liquid-gas transition. Therefore, a Van der Waals-like equation of state is expected to describe the different phases. Figures \ref{fig:VdW3} and \ref{fig:VdW4} are the analog of plotting `isotherms' in a ``P-V" diagram for a liquid-gas system. A drastic change in the derivatives $\partial\mu/\partial Q$ is observed when crossing the critical line. Thus,  we could identify the  different regions as a `liquid' phase, in which the black holes are highly sensitive to any change in the charges, and a `gaseous' phase, characterized by considerably smaller values of $\partial\mu/\partial Q$.

It is important to notice that the phase transitions always occur between black hole states with $Q_4\neq0$, as seen explicitly in the ``$\mu_4$-$Q_4$'' diagram (figure \ref{fig:VdW4}). This fact guarantees that the solutions between which the system is transitioning have the same asymptotic (UV, far from the horizon) structure \footnote{We thank one of the referees for pointing out this issue.}; it is the spin of the highest spin charge that sets the value of the $AdS$ radius and the central charge. As argued in \citep{Ferlaino:2013vga}, the corresponding asymptotic symmetry algebra for $N=4$ is a $\mathcal{W}$-algebra associated to the $(2,1,1)$ non-principal embedding of $sl(2)$ in $sl(4)$.
 
To conclude, in this paper we have studied the thermodynamic phase space properties of $sl(N;\mathds{R})$ higher spin black holes. We have identified the even-$N$ theories as those having well-defined (finite and stable) partition functions in the ensemble under consideration. These theories exhibit Hawking-Page transitions, just like any black hole coupled to a cosmological constant. Moreover, we find first order phase transitions between different higher spin black holes, as well as a second order transition and a critical point. According to the AdS/CFT correspondence, the phenomena observed in the bulk should have a counterpart at the boundary CFT. We leave the investigation of this issue for future work.     

\begin{acknowledgments}
MB was partially funded by Fondecyt \# 1141221. MB would also like thank F. Quevedo for hospitality at ICTP, where part of this work was done. AF acknowledges support from Fondecyt \# 1160282. GD acknowledges support from Fondecyt \# 1150463. The work of IR was partially funded by CONICYT PCHA/Doctorado Nacional \# 2015149744.
\end{acknowledgments}

\bibliography{Refs}
\bibliographystyle{ieeetr}
\end{document}